\documentclass[10pt, twocolumn]{IEEEtran}

\usepackage{graphicx}
\usepackage{psfrag}
\usepackage{subfigure}
\usepackage{epsfig}
\usepackage{amssymb}
\usepackage{amsmath}
\usepackage{subfigure}
\usepackage{bm}
\usepackage{algorithmic, algorithm}
\usepackage[left=0.625in, right=0.625in, top=0.75in, bottom=1in]{geometry}

\newcommand{\E}{{\mathbb E}}

\newcommand{\pr}{{\bf P}}

\newcommand{\tf}{{\tilde f}}
\newcommand{\tdelta}{{\tilde \delta}}

\newcommand{\tlambda}{{\tilde \lambda}}
\newcommand{\tLambda}{{\tilde \Lambda}}
\newcommand{\ttau}{{\tilde \tau}}

\newcommand{\tpr}{{\tilde {\bf \pr}}}
\newcommand{\tD}{{\tilde {D}}}
\newcommand{\tC}{{\tilde {C}}}
\newcommand{\tS}{{\tilde {S}}}

\newcommand{\pfa}{\text{PFA}}
\newcommand{\arl}{\text{ARL}}

\newcommand{\add}{\text{ADD}}
\newcommand{\esadd}{\text{ESADD}}

\newcommand{\ess}{\text{ess}}

\newtheorem{theorem}{Theorem}
\newtheorem{lemma}{Lemma}

\title{Quickest Change Detection with Mismatched Post-Change Models}
\author{Jingxian Wu, \IEEEmembership{Senior Member, IEEE,} and Jing Yang, \IEEEmembership{Member, IEEE},
\thanks{The authors are with the Department of Electrical Engineering,
University of Arkansas, Fayetteville, AR, 72701, U.S.A.}
\thanks{The work was supported in part by the U.S. National Science Foundation (NSF) under 
Grants ECCS-1202075, ECCS-1405403, and ECCS-1454471.}}

\begin{document}

\maketitle

\begin{abstract}
In this paper, we study the quickest change detection with mismatched post-change models. 
A change point is the time instant at which the distribution of a random process changes. 
The objective of quickest change detection is to minimize the detection delay of an unknown change point 
under certain performance constraints, such as average run length (ARL)
to false alarm or probability of false alarm (PFA). Most existing change detection procedures assume
perfect knowledge of the random process distributions before and after the change point. However, in many practical
applications such as anomaly detection, the post-change distribution is often unknown and needs to be estimated
with a limited number of samples. 
In this paper, 
we study the case that there is a mismatch between the true post-change distribution and the one used during detection. We analytically 
identify the impacts of mismatched post-change models on two classical detection procedures, the cumulative sum
(CUSUM) procedure and the Shiryaev-Roberts (SR) procedure. The impacts of mismatched models are characterized in terms of various finite
or asymptotic performance bounds on ARL, PFA, and average detection delay (ADD). It is shown that post-change model
mismatch results in an increase in ADD, and the rate of performance degradation depends on the difference between
two Kullback-Leibler (KL) divergences, one is between
the priori- and post-change distributions, and the other one is between the true and mismatched post-change distributions. 
\end{abstract}

\section{Introduction}
Change detection is the process of identifying the time instants at which the distribution
of a random process changes. It has a wide range of applications in various science, engineering, and financial fields, 
such as intrusion detection, anomaly detection, quality control, financial market analysis, and medical diagnosis, etc. 

Change detection methods can be classified into two categories, offline and online change point detections. In offline change 
detection, the detector estimates the locations of one or more change points based on the observations of the
entire random process or time sequence \cite{Zhang15}. Offline methods usually need to detect the number of change points before identifying
the location of each change point. Online change detection uses sequential analysis to detect whether a change point
has happened before the current time by using all currently observed samples \cite{Shiryaev61}-\cite{Banerjee15}. Online change detection usually needs to make
tradeoff among various performance metrics, such as detection delay, probability of false alarm (PFA), and average run length (ARL)
to false alarm, etc. 

Quickest change detection is an online detection method, and it 
aims at minimizing the detection delay of a change point under the constraints of an upper bound on PFA or a lower bound on ARL. 
The change point itself can be modeled as a random variable with prior distributions. If the prior distribution of the 
change point is known, then Bayesian change detection, such as the well known Shiryaev procedure \cite{Shiryaev61, Shiryaev63}, 
can be performed. 
In \cite{Tartakovsky05}, Tartakovsky and Veeravalli asymptotically characterize the moments of the 
detection delay of the Shiryaev procedure by letting the PFA goes to zero, and they show that the Shiryaev procedure
is asymptotically optimum in the Bayesian setting under some mild conditions. When the prior distribution 
of the change point is not known, the online change detection can be performed under the minimax criterion, that is, minimizing
the expected delay for some worst case change point distribution. One of the most commonly used minimax change detection procedures
is the cumulative sum (CUSUM) procedure proposed by Page \cite{Page54}. The asymptotic behavior of the CUSUM procedure 
are characterized
by Lorden \cite{Lorden71} for independently and identically distributed (i.i.d.) samples, and later extended by Lai \cite{Lai98} 
for non-i.i.d. samples. It is shown that the CUSUM procedure can minimize the worst-worst-case detection delay as the ARL lower bound 
goes to infinity. Another popular minimax change detection method is the Shiryaev-Roberts (SR) procedure \cite{Shiryaev61, Shiryaev63, 
Roberts66}. The asymptotic optimality
of the SR procedure are discussed in \cite{Pollak09} and \cite{Tartakovsky10}.

All above mentioned procedures require precise knowledge of the distribution functions before and after the change point. 
In many practical applications, such as anomaly detection, it is relatively easy to learn and estimate the prior-change distribution,
because there is usually a large
amount of data available before the change point, e.g., data collected through normal operation conditions.
On the other hand, it is usually
difficult to obtain an accurate estimate of the post-change distribution, especially for quickest change detection where a decision
needs to be made as soon as possible with a limited number of observations from the post-change distribution. In \cite{Lai98}, a modified
generalized likelihood ratio (GLR) test is developed to take into consideration of some unknown parameters in the post-change distribution,
and it is shown that the modified procedure can attain the same asymptotic lower bound of detection delay 
as the case of known post-change distribution. In \cite{Banerjee15}, a non-parametric quickest detection method that does 
not require prior knowledge of the distributions is proposed.

In this paper, we study the performance of quickest change detection with mismatched post-change distribution models. That is, there 
is a mismatch between the true post-change distribution and the one used in the detection procedure, while the 
detector is assumed to have ideal knowledge of the prior-change distribution. The mismatch can be caused by
the limited amount of training data after the change point. Specifically, we study the impacts 
of mismatched post-detection models on two classical minimax detection procedures, the CUSUM and SR procedures. The performance
of CUSUM and SR procedures with mismatched models is characterized by deriving various finite or asymptotic bounds on 
the PFA, ARL, and average detection delay (ADD). It is shown that the PFA and ARL of the procedures with 
mismatched post-detection model can attain the same bounds as those with ideal post-detection models. 
On the other hand, under the same ARL or PFA constraints, post-change model mismatch results in degradation of ADD, and the
rate of degradation is determined by the difference between the true and mismatched post-change distributions, which can be 
measured as the Kullback-Leibler divergence between the two distributions.

\section{Problem Formulation}

Consider two continuous functions $f(x)$ and $g(x)$ where $\lim_{x\to x_0} f(x) = \lim_{x\to x_0} g(x) = \infty$. We have
the following notations.
\begin{align}
f(x) \underset{x \to x_0}{\preceq} g(x) \Longleftrightarrow \lim_{x\to x_0} \frac{f(x)}{g(x)} \leq 1
\end{align}

If both $f(x) \underset{x \to x_0}{\preceq} g(x)$ and $g(x) \underset{x \to x_0}{\preceq} f(x)$,
then the two functions are called asymptotically equivalent as $x \to x_0$, and it is denoted as
\begin{align}
f(x) \underset{x \to x_0}{\asymp} g(x) \Longleftrightarrow \lim_{x\to x_0} \frac{f(x)}{g(x)} = 1
\end{align}

\subsection{System Model}
Consider a random process $x_1, x_2, \cdots $. Define ${\bf x}^{k:n} = [x_k, \cdots, x_n]$. 
Let ${\cal F}_x^n = \sigma({\bf x}^{1:n})$ be the sigma algebra generated by 
${\bf x}^{1:n}$.

Assume there is an unknown change point $\theta$, such that the distribution of the random process 
before the change point differs from that after the change point. 
Let $\pr_k$ and $\E_k$ denote the probability measure and the corresponding expectation when the change occurs
at $\theta = k$. Under $\pr_k$, the conditional probability density function (pdf) is $f_{0,n}(x_n|{\bf x}^{1:n-1})$
for $n < k$, and it is $f_{1,n}(x_n|{\bf x}^{1:n-1})$ for $n \geq k$. With such a notation, 
$\pr_\infty$ and $\E_\infty$ can be used to represent 
the probability measure and the corresponding expectation before the change point, that is, the change point happens
at $\theta = \infty$. 

Assume the change point is random and it 
follows a prior distribution $\pr(\theta=k) = \pi_k$, for $k = 1, \cdots, n$. 
Define the average probability measure $\pr({\cal E}) = \sum_{k=1}^{\infty} \pi_k \pr_k({\cal E})$,
and $\E$ is the expectation with respect to $\pr$.

Define the likelihood ratio of the samples ${\bf x}^{k:n}$ as
\begin{align} \label{eqn:lambda}
\Lambda_{k:n} = \prod_{i=k}^n \frac{ f_{1,i}(x_i|{\bf x}^{1:i-1})}{ f_{0,i}(x_i|{\bf x}^{1:i-1})} = \prod_{i=k}^n \lambda_i
\end{align}
where
\begin{align}
\lambda_i = \frac{f_{1,i}(x_i|{\bf x}^{1:i-1})}{f_{0,i}(x_i|{\bf x}^{1:i-1})}
\end{align}

It is assumed that $\frac{1}{n} \log \Lambda_{k:k+n}$ converges in probability $\pr_k$ to a constant, $D_{10}$, as 
$n \to \infty$. That is,
\begin{align} \label{eqn:convergence_10}
\lim_{n \to \infty}\pr_k\left( \left| \frac{1}{n} \log \Lambda_{k:k+n} - D_{10}\right| > \epsilon \right) = 0, \forall \epsilon > 0.
\end{align}
When the samples are independent, $D_{10}$ is the Kullback-Leibler divergence between the distributions $f_{1,n}$ and 
$f_{0,n}$.

\subsection{Detection Procedures}

The quickest change detection is performed sequentially by using the observed data sequence. Define 
a detection procedure $\delta$ as a mapping from the observed sequence ${\cal F}_x^n$ to a positive integer $k \leq n$
\begin{align}
\delta: {\cal F}_x^n \rightarrow \{k: k \leq n \},  n = 1, 2, \cdots
\end{align}
Since $\delta({\bf x}^{1:n}) \leq n$, $\delta({\bf x}^{1:n})$ is a stopping time.

Denote the change point detected by $\delta$ as $\tau$, then the PFA associated with method $\delta$ is defined as
\begin{align}
\pfa(\delta) = \pr_{\infty}(\tau < \theta)
\end{align}

The corresponding ARL is defined as
\begin{align}
\arl(\delta) = \E_{\infty}(\tau)
\end{align}

The average detection delay (ADD) associated with the method $\delta$ is defined as 
\begin{align}
\add(\delta) = \E\left[\tau-\theta | \tau \geq \theta \right] = \sum_{k=1}^{\infty} \pi_k\E_k\left[\tau-k | \tau \geq k \right]
\end{align}

We will study the performance of two classical minimax procedures: the CUSUM procedure 
and the SR procedure.

\subsubsection{CUSUM procedure}
The CUSUM procedure is
\begin{align}
\delta_c(A) = \inf\{n: C_n \geq A\}
\end{align}
where 
\begin{align}
C_n = \max_{1\leq k \leq n} \Lambda_{k:n}.
\end{align}
We set $\inf\{\emptyset\} = \infty$.

The test statistics $C_n$ can be recursively calculated as
\begin{align}
C_{n+1} = \max(1, C_{n}) \lambda_{n+1}, ~~~n \geq 1
\end{align}
with $C_0 = 0$.

It was shown by Lorden \cite{Lorden71}
that under the constraint that 
the ARL is greater than a threshold $\gamma$, as $\gamma \to \infty$, the CUSUM minimzes the ``worst-worst-case" 
detection delay defined as
\begin{align}
\esadd(\delta) = \sup_{\theta\geq 0}\left\{\ess\sup \E_\theta\left((\tau-\theta)^+|{\cal F}_x^\theta\right) \right\}
\end{align}
The result was generalized by Lai \cite{Lai98} to systems with non-i.i.d. samples.

\subsubsection{SR procedure}
The SR procedure is
\begin{align}
\delta_s(A) = \inf\{n: S_n \geq A\}
\end{align}
where 
\begin{align}
S_n = \sum_{k=1}^n \Lambda_{k:n}.
\end{align}

The test statistics $S_n$ can be recursively calculated as
\begin{align}
S_{n+1} = (1 + S_n) \lambda_{n+1}, ~~~n \geq 1
\end{align}
with $S_0 = 0$.


Under the constraint that the ARL is greater than a threshold $\gamma > 1$, it is shown by Pollak and Tartakovsky in \cite{Pollak09} 
that the SR procedure can minimize the following metric
\begin{align}
\text{RIADD} = \frac{\sum_{k=0}^{\infty}\E_k(\tau-k)^+}{\E_{\infty}(\tau)}
\end{align}

The asymptotic ADD of both CUSUM and SR procedures are studied in \cite{Tartakovsky05}. It is shown that
if the convergence condition in \eqref{eqn:convergence_10} is satisfied,
then 
\begin{align}
\pfa(\delta_c(A)) &\leq \alpha \\
\pfa(\delta_s(A)) &\leq \alpha
\end{align}
for $A = {\bar \theta}/\alpha$, where $\bar \theta = \sum_{k=1}^{\infty} k \pi_k$ is the prior mean of the change point.

In addition, for $A = {\bar \theta}/\alpha$, 
\begin{align}
\add(\delta_c(A)) &\underset{\alpha\to 0}{\asymp} \frac{|\log \alpha|}{D_{10}} \\
\add(\delta_s(A)) &\underset{\alpha\to 0}{\asymp} \frac{|\log \alpha|}{D_{10}} 
\end{align}

\subsection{Detection Procedures with Mismatched Models}

The above detection procedures require the knowledge of the distributions of $x$ before and after
the change point.
In this paper we will
consider the model mismatch case that $f_{0,n}(x)$ is perfectly known, yet there are mismatches for 
the post-change distribution
$f_{1,n}(x)$. Denote $\tf_{1,n}(x)$ as the model used by the detection method, and 
$f_{1,n}(x)$ as the true model. We will study how the post-change model mismatch will affect the performance 
of the CUSUM and SR detection procedures.

With the mismatched model $\tf_{1,n}(x)$, define the mismatched likelihood ratio
\begin{align} \label{eqn:lambda_t}
\tlambda_i = \frac{\tf_{1,i}(x_i|{\bf x}^{1:i-1})}{f_{0,i}(x_i|{\bf x}^{1:i-1})},
\end{align}
and 
\begin{align} \label{eqn:tlambda}
\tLambda_{k:n} = \prod_{i=k}^n \tlambda_i.
\end{align}

Let $\tpr_k$ denote the mismatched probability measure such that under $\tpr_k$, 
the conditional probability density function (pdf) is $f_{0,n}(x_n|{\bf x}^{1:n-1})$
for $n < k$, and it is $\tf_{1,n}(x_n|{\bf x}^{1:n-1})$ for $n \geq k$.


The corresponding mismatched test statistics for the CUSUM and SR procedures can be written, respectively, as
\begin{align}
\tC_n &= \max_{1 \leq k \leq n} \tLambda_{k:n} \label{eqn:tc_n}\\
\tS_n &= \sum_{k=1}^n \tLambda_{k:n} \label{eqn:ts_n}
\end{align}

The above test statistics can be calculated recursively as
\begin{align} 
\tC_{n+1} &= \max(1, \tC_n)\tlambda_{n+1} \label{eqn:tc_n+1} \\
\tS_{n+1} &= (1 + \tS_n) \tlambda_{n+1} \label{eqn:ts_n+1}
\end{align}

The CUSUM and SR procedures with mismatched models can be represented, respectively, as
\begin{align}
\tdelta_c(A) &= \ttau_c = \inf\{n:\tC_n \geq A\} \label{eqn:ttauc} \\
\tdelta_s(A) &= \ttau_s = \inf\{n:\tS_n \geq A\} \label{eqn:ttaus}
\end{align}

\section{Impacts of Model Mismatch on ARL and PFA}
In this section, we study the impacts of post-change model mismatch of the performance of the CUSUM and SR detection
procedures, in terms of the ARL and the PFA.

\subsection{ARL}

The ARLs of the CUSUM and SR procedures with mismatched post-change models are studied in this subsection.

\begin{lemma}
$\tS_n-n$ is a martingale under the probability measure $\pr_{\infty}$,
and $\E_\infty(\tS_n - n)=0$. 
\end{lemma}
\begin{IEEEproof}
Under the probability measure $\pr_{\infty}$, we have 
\begin{align}
\E_{\infty}(\tLambda_{k:n}) = \int \frac{d \tpr_k({\bf x}^{k:n}|{\bf x}^{1:k-1})}{d \pr_0({\bf x}^{k:n}|{\bf x}^{1:k-1})} d \pr_0({\bf x}^{k:n}|{\bf x}^{1:k-1}) = 1
\end{align}

For the SR procedure, based on the recursive calculation of $\tS_{n+1}$, we have
\begin{align} \label{eqn:Ets_n1}
\E_{\infty}(\tS_{n+1}|\tS_n) = 1+ \tS_n
\end{align}
Thus $\tS_{n}-n$ is a Martingale. 

From the definition of $\tS_n$, we have $\E_{\infty}(\tS_n) = \sum_{k=1}^n \E_{\infty}(\tLambda_{k:n}) = n$.
\end{IEEEproof}

\begin{lemma}
The ARL for both the CUSUM and SR procedures with mismatched post-change models satisfy
\begin{align}
\arl(\tdelta_c(A)) &\geq A \\
\arl(\tdelta_s(A)) &\geq A
\end{align}
\end{lemma}
\begin{IEEEproof}
If $\E_\infty(\ttau_s) = \infty$, then $\arl(\tdelta_c(A)) = \infty \geq A$. 

We will next consider the case when $\E_\infty(\ttau_s) < \infty$. From 
\eqref{eqn:Ets_n1}, it is straightforward
that $\E_{\infty}[|\tS_{n+1}-(n+1) - \tS_{n}-n|\tS_n] = 0$.
Based on the optional stopping theorem, we have
\begin{align}
\E_{\infty}(\tS_{\ttau_s}-\ttau_s) = \E_{\infty}(\tS_1-1) = 0
\end{align}
Thus
\begin{align}
\E_{\infty}(\ttau_s) = \E_{\infty}(\tS_{\ttau_s}) \geq A
\end{align}

Since $\tC_n \leq \tS_n$, under the same threshold $A$, we have $\ttau_c \geq \ttau_s$, 
thus $\E_{\infty}(\ttau_c) \geq \E_{\infty}(\ttau_s) \geq A$.
\end{IEEEproof}

\subsection{PFA}

\begin{lemma}
The PFA of the SR procedure with mismatched model is upper bounded by 
\begin{align}
\pfa(\tdelta_s(A)) \leq \min\left\{\frac{\bar \theta}{A}, 1\right\}
\end{align}
where ${\bar \theta} = \sum_{k=1}^{\infty} k \pi_k$ is the priori mean of the change point $\theta$.
\end{lemma}
\begin{IEEEproof}
Since $\tS_n-n$ is a martingale with respect to $\pr_{\infty}$, $\tS_n$ is a sub-martingale with respect
to $\pr_{\infty}$. Based on Doob's inequality, we have
\begin{align}
\pr_{\infty}(\ttau_s < n) = \pr_{\infty}\left( \max_{1 \leq k < n} \tS_k \geq A\right) \leq \frac{n}{A}
\end{align}
Therefore 
\begin{align}
\pr(\ttau_s < \theta) = \sum_{k=1}^{\infty} \pi_k \pr_{\infty}(\ttau_s < k) = \frac{\bar \theta}{A}
\end{align}
\end{IEEEproof}

\begin{lemma}
The PFA of the CUSUM procedure with mismatched model is upper bounded by 
\begin{align} \label{eqn:pfa_cusum}
\pfa(\tdelta_c(A)) \leq \frac{1}{A} \leq \frac{\bar \theta}{A}
\end{align}
\end{lemma}
\begin{IEEEproof}
It can be easily shown that $\tC_n$ is a sub-Martingale because 
\begin{align}
\E_{\infty}(\tC_{n+1}|\tC_n) = \max(1, \tC_n) \geq \tC_n
\end{align}
In addition, $\E_{\infty}(\tC_{n}) = \max_{1 \leq k \leq n}\E_{\infty}(\Lambda_{k:n}) = 1$.

Based on Doob's inequality, we have 
\begin{align}
\pr_{\infty}(\ttau_c < n) = \pr_{\infty}\left( \max_{1 \leq k < n} \tC_k \geq A\right) \leq \frac{1}{A}
\end{align}
Therefore 
\begin{align}
\pr(\ttau_c < \theta) = \sum_{k=1}^{\infty} \pi_k \pr_{\infty}(\ttau_c < k) = \frac{1}{A}
\end{align}

The second inequality in \eqref{eqn:pfa_cusum} is from the fact that $\bar \theta \geq \min(\theta) = 1$.
\end{IEEEproof}

From the above results, we have
\begin{align} 
\pfa\left(\delta_c\left(\frac{1}{\alpha}\right) \right) &\leq \alpha \label{eqn:pfa_c} \\
\pfa\left(\delta_s\left(\frac{\bar \theta}{\alpha}\right) \right) & \leq \alpha \label{eqn:pfa_s}
\end{align}

Based on the above analysis, it can be seen that a mismatch in the post-change distribution have no impact on the ARL lower bound 
or PFA upper bound, because the ARL and PFA are calculated with respect to the probability measure $\pr_{\infty}$, and they
will only be affected by the distribution prior to the change. 

\section{ADD with Mismatched Models}

The ADD of CUSUM and SR procedures with mismatched post-change distributions are studied in this section.

In order to study the impact of model mismatch on ADD, we define the likelihood ratio between the true and mismatched
post-change distributions as
\begin{align}
\tlambda^{11}_n = \frac{f_1(x_n|{\bf x}^{1:n-1})}{\tf_1(x_n|{\bf x}^{1:n-1})}
\end{align}
and
\begin{align} \label{eqn:tlambda_11}
\tLambda^{11}_{k:n} = \prod_{i=k}^n \tlambda^{11}_i
\end{align}

In addition to the convergence assumption in \eqref{eqn:convergence_10}, it is assumed
that $\frac{1}{n} \log \tLambda^{11}_{k:k+n}$ converges in probability $\pr_k$ to a constant, $\tD_{11}$ , that is
\begin{align} \label{eqn:convergence_t11}
\lim_{n \to \infty}\pr_k\left( \left| \frac{1}{n} \log \tLambda^{11}_{k:k+n} - \tD_{11}\right| > \epsilon \right) = 0, \forall \epsilon > 0,
\end{align}

We will study the ADD by considering two cases: $D_{10}-\tD_{11} > 0$ or $D_{10}-\tD_{11} < 0$. 

\subsection{$D_{10}-\tD_{11} > 0$}
We will derive an asymptotic upper bound on ADD as the PFA $\alpha \to 0$. 
To obtain the upper bound, define a new stopping time
\begin{align} \label{eqn:beta_def}
\beta(A) = \inf\{n \geq k:  \tLambda_{k:n} \geq A\}
\end{align}
We have the following lemma regarding the asymptotic behavior of $\beta(A)$ as $A \to \infty$.

\begin{lemma} \label{lem:upper_qm}
Assume the convergence condition in \eqref{eqn:convergence_10} and \eqref{eqn:convergence_t11} are satisified. If $\tD_{11} < D_{10}$,
as $A \to \infty$, we have
\begin{align}
\E_k[(\beta(A)-k)^+]  \underset{A\to \infty}{\preceq}  \frac{\log A}{D_{10}-\tD_{11}}, ~~~\forall q \neq m
\end{align}
where $a^+ = a$ if $a \geq 0$ and $a^+ = 0$ otherwise.
\end{lemma}
\begin{IEEEproof}
From \eqref{eqn:lambda}, \eqref{eqn:tlambda} and \eqref{eqn:tlambda_11}, we have
\begin{align} \label{eqn:log_split}
\log \tLambda_{k:n} = \log \Lambda_{k:n} - \log \tLambda_{k:n}^{11}
\end{align}

Based on the convergence condition in \eqref{eqn:convergence_10} and \eqref{eqn:convergence_t11}, for any $\epsilon > 0$, 
there exists $N_{\epsilon} < \infty$ such that for all $n > N_{\epsilon}$, 
\begin{align}
\left|\frac{\log \Lambda_{k:n}}{n-k+1} - D_{10}\right| < \frac{\epsilon}{2}, ~~a.s. \\
\left|\frac{\log \tLambda_{k:n}^{11}}{n-k+1} - \tD_{11}\right| < \frac{\epsilon}{2}, ~~a.s.
\end{align}
with respect to $\pr_k$.

Thus from \eqref{eqn:log_split}
\begin{multline}
\left|\frac{\log \tLambda_{k:n}}{n-k+1}  - (D_{10}-\tD_{11}) \right|  \leq \\
 \left|\frac{\log \Lambda_{k:n}}{n-k+1} - D_{10}\right| + 
\left|\frac{\log \tLambda_{k:n}^{11}}{n-k+1} - \tD_{11}\right| < \epsilon, ~~a.s.
\end{multline}
for all $n > N_{\epsilon}$ with respect to $\pr_k$. 
 
For any $0 < \epsilon < D_{10}-\tD_{11}$, define 
\begin{align}
T_{\epsilon} = \sup\left\{n \geq 1: \left|\frac{\log \tLambda_{k:n}}{n-k+1}  - (D_{10}-\tD_{11}) \right| \geq \epsilon \right\}
\end{align}
Thus $T_{\epsilon} \leq N_{\epsilon} < \infty$.

Based on the definition of $\beta(A)$ in \eqref{eqn:beta_def}, it is obvious that
\begin{align}
\tLambda_{k:\beta(A)-1} < A
\end{align}

When $\beta(A) -1  > T_{\epsilon}$, we have 
\begin{align}
\frac{\log\tLambda_{k:\beta(A)-1}}{\beta(A)-k} -(D_{10}-\tD_{11})
   > -\epsilon
\end{align}
Thus 
\begin{align*}
\beta(A) &< k + \frac{\log\tLambda_{k:\beta(A)-1}}{D_{10}-\tD_{11}-\epsilon}, \text{~~when~~} \beta(A) > T_{\epsilon}+1 \\
\beta(A) &\leq T_{\epsilon}+1, \text{~~when~~} \beta(A) \leq T_{\epsilon}+1 
\end{align*}

Therefore
\begin{align}
\beta(A) &< k+1+\frac{\log\tLambda_{k:\beta(A)-1}}{D_{10}-\tD_{11}-\epsilon}+T_{\epsilon} \\
&< k+1+\frac{\log A}{D_{10}-\tD_{11}-\epsilon}+T_{\epsilon} 
\end{align}

Since $\epsilon$ can be arbitrarily small and $T_{\epsilon}< \infty$, let $\epsilon \to 0$ we have
\begin{align}
\lim_{A \to \infty} \frac{\E_k[\beta(A)-k] }{\frac{\log A}{D_{10}-\tD_{11}}} < 1
\end{align}
\end{IEEEproof}

With the results in Lemma \ref{lem:upper_qm}, we can obtain an asymptotic upper bound of the ADD with
mismatched post-change models, and the results are given in the following theorem.

\begin{theorem} \label{thm:add_upper}
Assume the convergence condition in \eqref{eqn:convergence_10} and \eqref{eqn:convergence_t11} are satisified. 
Let $\pfa < \alpha$. If $\tD_{11} < D_{10}$,
as $\alpha \to 0$, we have
\begin{align}
\E[\ttau_c-\theta|\ttau_c > \theta ] &\underset{\alpha \to 0}{\preceq} \frac{\frac{|\log \alpha|}{1-\alpha}}{D_{10}-\tD_{11}} \\
\E[\ttau_s-\theta|\ttau_s > \theta ] &\underset{\alpha \to 0}{\preceq} \frac{\frac{\log{\bar \theta}-\log \alpha}{1-\alpha}}{D_{10}-\tD_{11}} 
\end{align}
\end{theorem}
\begin{IEEEproof}
The ADD of the CUSUM procedure with mismatched model can be alternatively written as
\begin{align} \label{eqn:add_c}
\E[\ttau_c-\theta|\ttau_c > \theta ] = \frac{1}{\pr_\infty(\ttau_c \geq \theta)}\sum_{k=1}^\infty \pi_k {\E_k(\ttau_c-k)^+}
\end{align}
By definition, we have $\ttau_c < \beta(A)$, thus from Lemma \ref{lem:upper_qm}, 
\begin{align} \label{eqn:tau_k_c}
\E_k(\ttau_c-k)^+ \underset{A\to \infty}{\preceq} \frac{\log A}{D_{10}-\tD_{11}} 
\end{align}

From \eqref{eqn:pfa_c}, we can set $A = \frac{1}{\alpha}$ to guarantee $\pfa < \alpha$. 
Thus $\pr_\infty(\ttau_c \geq \theta) = 1-\pfa \geq 1 - \alpha$. 
Combining \eqref{eqn:add_c}, \eqref{eqn:tau_k_c} and the above results, we have
\begin{align}
\E[\ttau_c-\theta|\ttau_c > \theta ] \underset{\alpha\to 0}{\preceq} \frac{\frac{|\log \alpha|}{1-\alpha}}{D_{10}-\tD_{11}} 
\end{align}

For the upper bound of $\ttau_s$, from \eqref{eqn:pfa_s}, we can set $A = \frac{\bar \theta}{\alpha}$ to ensure $\pfa < \alpha$. 
The remaining procedures are the same as the analysis of $\ttau_c$. 
\end{IEEEproof}

\subsection{$D_{10}-\tD_{11} < 0$}

To facilitate analysis, define a new stopping time
\begin{align} \label{eqn:beta_def}
\zeta_v(A) = \inf\{n \geq v:  \tLambda_{v:n} \geq A\}
\end{align}
We have the following lemma regarding the behavior of $\zeta_v(A)$ when $D_{10}-\tD_{11} < 0$. 


\begin{lemma} \label{lem:infty}
Assume the convergence condition in \eqref{eqn:convergence_10} and \eqref{eqn:convergence_t11} are satisified. 
If $D_{10}-\tD_{11} < 0$ and $A > 1$, then
\begin{align}
\E_k[(\zeta_v(A)-k)^+] = \infty
\end{align}
\end{lemma}
\begin{IEEEproof}
Define $Z_{n} = \log \tLambda_{v:n} = Z_{n-1} + \log \tlambda_n$.

1) When $v < k$, we have
\begin{align}
\E_k(Z_{n+1} | Z_n) = Z_{n} + (D_{10}-\tD_{11}), \text{~for all~} n \geq k
\end{align}
Thus $Z_{n} - n(D_{10}-\tD_{11})$ forms a martingale for all $n \geq k$, with
$\E_k(Z_k) = D_{10}-\tD_{11}$

Proof by contradiction. Assume $E_k[\zeta_v(A)| \zeta_v(A)>k] < \infty$. Then based on optional stopping theorem,
we have
\begin{align}
&\E_k[Z_{\zeta_v(A)}| \zeta_v(A)>k] - \E_k[\zeta_v(A)| \zeta_v(A)\geq k](D_{10}-\tD_{11}) = \nonumber \\
& \E_k[Z_k] - k(D_{10}-\tD_{11})  = -(k-1)(D_{10}-\tD_{11})
\end{align}

Thus when $A > 1$, 
\begin{align}
\E_k[\zeta_v(A)-(k-1)|\zeta_v(A)\geq k] &= -\frac{\E_k[Z_{\zeta_v(A)}]}{|D_{10}-\tD_{11}|} \\
&\leq -\frac{ \log A}{|D_{10}-\tD_{11}|}<0
\end{align}
This contradicts with $\E_k[\zeta_v(A)-(k-1)|\zeta_v(A)\geq k]\geq 0$. Thus $\E_k[\zeta_v(A)|\zeta_v(A)\geq k]= \infty$.
Since $\pr(\zeta_v(A)\geq k) > 0$, we have $\E_k[(\zeta_v(A)-k)^+] = \infty$. 

2) When $v \geq k$, we have
\begin{align}
\E_k(Z_{n+1} | Z_n) = Z_{n} + (D_{10}-\tD_{11}), \text{~for all~} n \geq v
\end{align}
Thus $Z_{n} - n(D_{10}-\tD_{11})$ forms a martingale for all $n \geq v$, with 
$E_k(Z_v) = D_{10}-\tD_{11}$. 

Proof by contradiction. Assume $E_k[\zeta_v(A)|\zeta_v(A)\geq v] < \infty$. Then based on optional stopping theorem,
we have
\begin{align}
&\E_k[Z_{\zeta_v(A)}|\zeta_v(A)\geq v] - \E_k[\zeta_v(A)|\zeta_v(A)\geq v](D_{10}-\tD_{11}) = \\
& \E_k[Z_v] - v(D_{10}-\tD_{11})  = -(v-1)(D_{10}-\tD_{11})
\end{align}

Thus when $A > 1$, 
\begin{align}
\E_k[\zeta_v(A)-(v-1)|\zeta_v(A)\geq v] &= -\frac{\E_k[Z_{\zeta_v(A)}]}{|D_{10}-\tD_{11}|} \\
&\leq -\frac{ \log A}{|D_{10}-\tD_{11}|}<0
\end{align}
This contradicts with $\E_k[\zeta_v(A)-(v-1)|\zeta_v(A)\geq v]\geq 0$. Thus $\E_k[(\zeta_v(A)-v)^+]= \infty$.
Since $v \geq k$, we have $\E_k[(\zeta_v(A)-k)^+] \geq \E_k[(\zeta_v(A)-v)^+] = \infty$.
\end{IEEEproof}

\begin{theorem} \label{thm:infty_add}
Assume the convergence condition in \eqref{eqn:convergence_10} and \eqref{eqn:convergence_t11} are satisified. 
Let $\pfa < \alpha < 1$. 
If $D_{10}-\tD_{11} < 0$, then the ADD of the CUSUM procedure satisfies
\begin{align}
\E[\ttau_c - \theta| \ttau_c \geq \theta] = \infty
\end{align}
\end{theorem}
\begin{IEEEproof}
Based on the definition of $\ttau_c$ and $\zeta_v(A)$, we have
\begin{align}
\ttau_c = \min_{v} \zeta_v(A)
\end{align}

From \eqref{eqn:pfa_c}, we can set $A = \frac{1}{\alpha}$ to ensure $\pfa < \alpha$. Since $\alpha < 1$, thus $A > 1$. 
From Lemma \ref{lem:infty}, we have
\begin{align}
\E_k(\ttau_c - k)^+ = \min_v \E(\zeta_v(A)-k)^+ = \infty
\end{align}
Thus
\begin{align}
\E[\ttau_c - \theta| \ttau_c \geq \theta] = \frac{1}{\pr(\ttau_c \geq \theta)} \sum_{k=1}^\infty \pi_k \E_k(\ttau_c - k)^+ = \infty
\end{align}
\end{IEEEproof}

Please note the infinity ADD result in Theorem \ref{thm:infty_add} is not an asymptotic result and it only requires $\alpha <1$. 
Such a non-asymptotic result is in general not true for the SR procedure. 
If the asymptotic condition $\log(\tS_n) \underset{\alpha \to 0}{\asymp}  \log(\tC_n)$ is satisfied, then
we have $\E[\ttau_s - \theta| \ttau_s \geq \theta] \underset{\alpha \to 0}\rightarrow \infty$ when $D_{10}-\tD_{11}<0$.

\section{Numerical Results}
Numerical and simulation results are provided in this section to verify the analytical bounds obtained in this
paper. In the simulations, all data follow a two-dimension multivariate Gaussian distribution with zero-mean and 
covariance matrix 
\begin{align}
{\bf R} = \left[\begin{array}{cc}
1 & \rho \\
\rho & 1
\end{array} \right].
\end{align} 
The coefficient $\rho$ before and after the change point is 0 and 0.5, respectively. The prior distribution
of the change point is the geometric distribution with parameter $p_0$, that is, $\pi_k = (1-p_0)^{k-1} p_0$.
We set $p_0 = 0.1$ in all simulations.

\begin{figure}
    \begin{center}
      \includegraphics[width=0.5\textwidth]{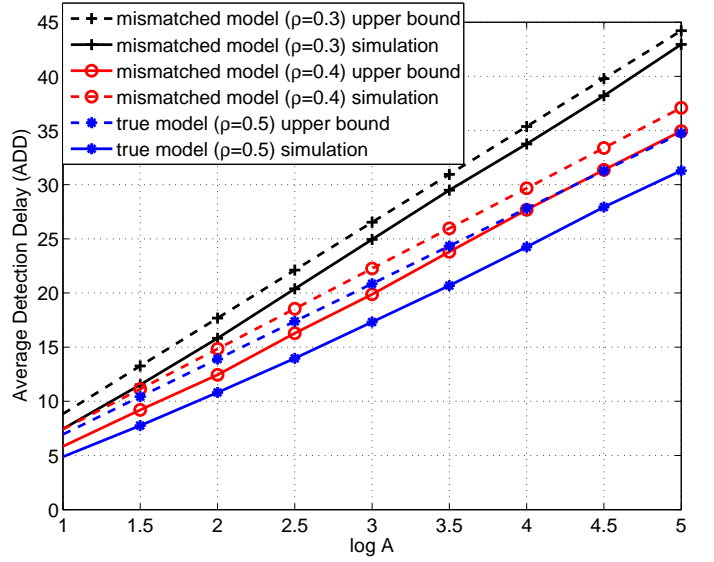}
     \caption{Average detection delay of the CUSUM procedure.} \label{fig:delay_cusum}
   \end{center}
 \end{figure}

Fig. \ref{fig:delay_cusum} compares the ADD of CUSUM with true or mismatched post-change point models as a function
of the logarithm of the detection threshold. As indicated by Theorem \ref{thm:add_upper}, the asymptotic upper bound is linear in $\log A$,
with slope inversely proportional to $D_{10}-\tD_{11}$. Under the configuration in this simulation, we have $D_{10} = 0.1438$,
$\tD_{11} = 0.0308$ for $\rho = 0.3$, and $\tD_{11} = 0.0090$ for $\rho = 0.4$. Thus the ADD of the true model has the
smallest slope, and the ADD of the mismatched model with $\rho = 0.3$ has the largest slope.
The ADDs obtained through simulations are also 
approximately linear in $\log A$, and they follow the same trends as their respective upper bounds.  

\begin{figure}
    \begin{center}
      \includegraphics[width=0.5\textwidth]{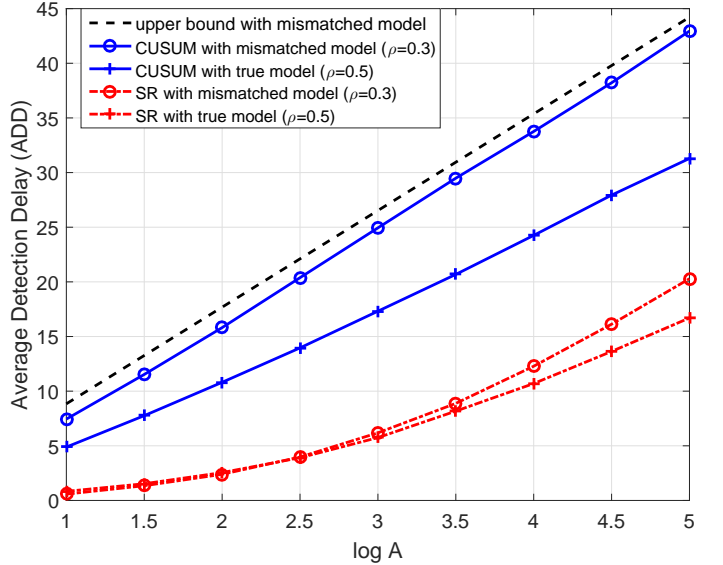}
     \caption{Average detection delay of both CUSUM and SR procedures.} \label{fig:delay_cusum_rs}
   \end{center}
 \end{figure}

\begin{figure}
    \begin{center}
      \includegraphics[width=0.5\textwidth]{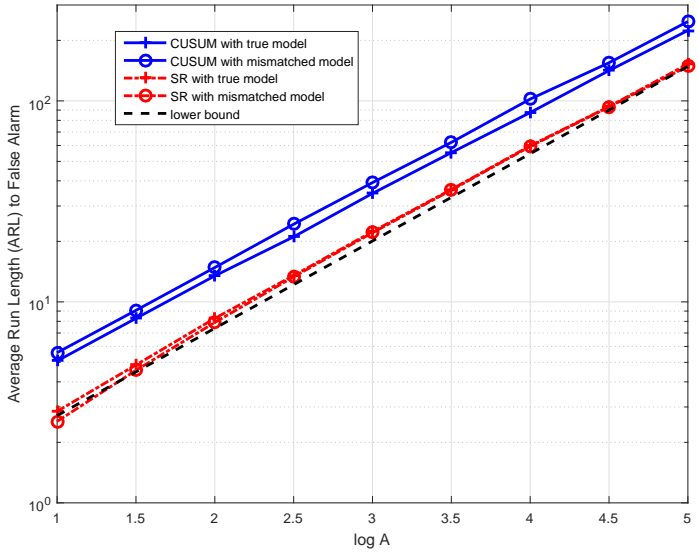}
     \caption{Average run length to false alarm.} \label{fig:arl}
   \end{center}
 \end{figure}

\begin{figure}
    \begin{center}
      \includegraphics[width=0.5\textwidth]{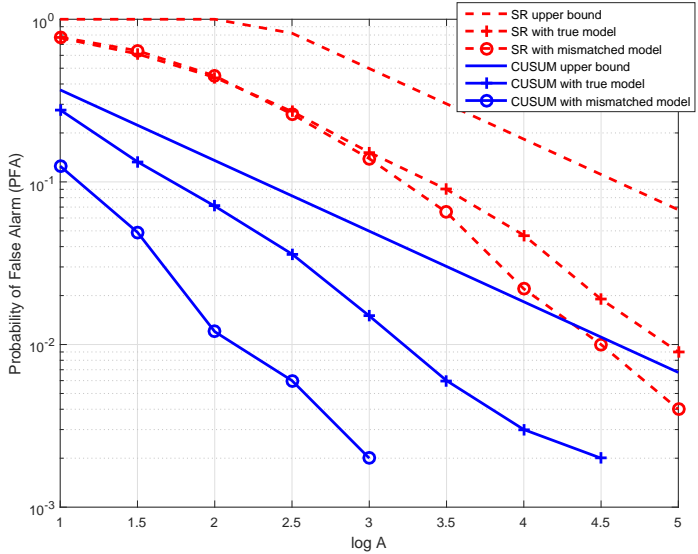}
     \caption{Probability of false alarm.} \label{fig:pfa}
   \end{center}
 \end{figure}

We compare the ADDs of systems with the CUSUM and SR procedures in Fig. \ref{fig:delay_cusum_rs}. For comparison purpose,
we use the same threshold $A$ for both procedures. It can be seen that 
even the upper bound is pretty tight for the CUSUM procedure, it is loose for the SR procedure. The SR procedure
considerably outperforms the CUSUM procedure in terms of ADD, for both true models and mismatched models. The performance
gain of SR in terms of ADD is achieved at the cost of PFA and ARL, as will be shown in Figs. \ref{fig:pfa} and \ref{fig:arl}.
When $\log A > 3.5$,
we can see that the ADD curves of both CUSUM and SR procedures with mismatched models 
share the same slope as the upper bound.

The ARLs obtained with various detection procedures are shown in Fig. \ref{fig:arl} as a function of the logrithm of the
threshold $A$. For the mismatched model
after the change point, the coefficient $\rho$ is assumed to be 0.3, while the true model uses $\rho =0.5$. 
The mismatched model has very small impact on the ARL to false alarm, for both CUSUM and SR procedures. 
For the CUSUM procedure, using $\rho = 0.3$ instead of its true value 0.5 results in a slight increase in the ARL.
For the SR procedure, the ARLs of system with true or mismatched post-change model are almost the same. 
The SR procedure has a smaller ARL than CUSUM.

Fig. \ref{fig:pfa} shows the PFA of various detection procedures and their corresponding upper bounds. All parameters
are the same as Fig. \ref{fig:arl}. 
The CUSUM procedure outperforms the SR procedure in terms of PFA, for both true and mismatched post-change model. 
It is interesting to note that using a mismatched coefficient of $\rho = 0.3$ leads to a smaller PFA than using
the true coefficient $\rho = 0.5$, for both CUSUM and SR procedures.

\section{Conclusions}

We have studied the quickest change detection when there is a mismatch between the true
post-change distribution and the one used by the detection procedures. The performance of two
commonly used minimax procedures, the CUSUM procedure and the Shiryaev-Roberts procedure, have been
characterized in this paper. The impacts of mismatched post-change model on the ARL, PFL, and ADD
have been identified in terms of various finite or asymptotic performance bounds. Detection procedures
with mismatched post-change models can attain the same ARL lower bound or PFA upper bound as those with
true models. On the other hand, the ADD will be increased
significantly due to model mismatch. When $D_{10} - \tD_{11} > 0$, the ADD is asymptotically upper bounded.
When $D_{10} - \tD_{11} < 0$, the ADD of the CUSUM procedure is infinity.

\end{document}